\documentclass[twocolumn,aps]{revtex4}
\usepackage{graphicx}
\usepackage{latexsym}
\usepackage{amssymb}
\usepackage{amsfonts}
\begin{document}
%
\title{Two--dimensional Copolymers and Multifractality:\\
 Comparing Perturbative Expansions, MC Simulations, and Exact Results}
\author{C.~von~Ferber}
\email[]{ferber@physik.uni-freiburg.de}
\affiliation{Theoretical Polymer Physics, Hermann-Herder-Str. 3, 79104
  Freiburg University, Germany}
\author{Yu.~Holovatch}
\email[]{hol@icmp.lviv.ua}
\affiliation{Institute for Condensed Matter
  Physics of the National Academy of Sciences of Ukraine, 79011 Lviv,
  Ukraine} \affiliation{Ivan Franko National University of Lviv, 79005
  Lviv, Ukraine}
\begin{abstract}
  We analyze the scaling laws for a set of two different species of
  long flexible polymer chains joined together at one of their
  extremities (copolymer stars) in space dimension $D=2$. We use a
  formerly constructed field--theoretic description and compare our
  perturbative results for the scaling exponents with recent
  conjectures for exact conformal scaling dimensions derived by a
  conformal invariance technique in the context of $D=2$ quantum
  gravity. A simple MC simulation brings about reasonable agreement
  with both approaches. We analyse the remarkable multifractal
  properties of the spectrum of scaling exponents.
\end{abstract}
\date{\today}
\maketitle
Polymers serve as a testing ground for a field theory since the early
seventies \cite{deGennes79,polymerRG}.  The path integral formulation
\cite{Kleinert95} allows for a direct interpretation of the paths as
conformations of random walks (RWs) and self avoiding walks (SAWs).
Closed paths in this theory have a multiplicity of the number of
components $M$ of the field. The formal limit $M=0$ excludes these
loops and yields the polymer limit of field theory, that describes
self and mutually interacting paths.
The field theoretical description of the scaling
properties of polymer chains in good solvents has been generalized to the
case of multicomponent polymer solutions \cite{ternary} and linked
polymers \cite{stars,stars2d}. Recently, the theory was suited to
describe polymer networks of different species
\cite{FerHol97,costars}. The simplest nontrivial case of a
heterogeneous polymer network -- a star-shaped copolymer in $D=2$
dimensions -- is the subject of the present report.

Polymer field theory usually has to be evaluated in terms of a
truncated perturbation theory series.  Obviously, it is always of
special importance to compare these with the exact results if such
exist. As a rule they are available for $D=1,2$. For the critical
exponents that describe the scaling of homogeneous polymer chains and
star polymers (or, more generally, homogeneous polymer networks) the
perturbative approach \cite{stars} is in fair agreement with the exact
data for $D=2$ \cite{stars2d}. For the heterogeneous case however,
only perturbative results \cite{FerHol97,costars} were available until
recently.  An exact solution for scaling exponents has been recently
proposed that uses methods of conformal invariance on random graphs
\cite{Duplantier99}. Below we undertake a comparative analysis of the
properties of the perturbative solution and of the exact one and
complement these results by MC simulations. We show where the
approaches agree and explain why they differ in general. In this
respect our paper confirms some of the speculative results of
\cite{Duplantier99} derived from the algebra of conformal dimensions
for copolymer stars on 2$D$ random graphs (quantum gravity).

A remarkable feature of the scaling spectrum of a copolymer star is that
it possesses multifractal properties that can also be analyzed
in terms of field theory and the conformal invariance approach.
\cite{FerHol97,costars,Duplantier99,Cates87,Hentschel83,Halsey86}.
\paragraph{Copolymer stars and their scaling exponents}\label{II}
Star polymers as the most simple nontrivial examples of polymer
networks \cite{stars,stars2d} are created by linking together the end
points of polymer chains at a common core. Let us now consider a
general star polymer made of chains of {\em two different species}
(while all chains have equilibrium size $R$). When such a {\em
  copolymer star} is immersed in a good solvent its asymptotic
properties are universal in the limit of long chains
\cite{deGennes79,polymerRG,Kleinert95}.  In particular, the number of
configurations (the partition function $Z_{*}$) scales with $R$ on
some scale $\ell$ as:
\begin{equation}\label{1}
Z_{*}\sim(R/\ell)^{-\lambda_{f_1,f_2}},
\end{equation}
where $\lambda_{f_1,f_2}$ constitute a family of copolymer star
exponents \cite{FerHol97,costars}. These are universal and depend only
on the space dimension $D$ and the number of chains of different
species 1 and 2 ($f_1$ and $f_2$ correspondingly). If a nonvanishing
mutual avoidance interaction is present between the chains of species
1 and 2 polymer field theory predicts two nontrivial physically
different regimes \cite{ternary} for the scaling behavior of true
copolymer stars described by Eq. (\ref{1}): In the first case the
polymers of species 1 behave as RWs and the polymers of species 2 as
SAWs, whereas in the second case both species display RW behavior
\cite{note1}. For these two relevant cases a renormalization group
description of the exponents $\lambda_{f_1,f_2}$ has been given in
terms of the scaling dimensions of appropriately composed composite
operators of polymer field theory \cite{FerHol97}. In particular, the
resulting perturbation theory was derived in the form of an
$\varepsilon=4-D$ expansion as well as evaluated directly for fixed
$D$ \cite{note2}. Following Ref. \cite{FerHol97} where the copolymer
star exponents $\eta_{f_1,f_2}$ were calculated, the
$\varepsilon$-expansion for the $\lambda$-exponents of a copolymer star
composed of two mutually avoiding sets of RWs
($\lambda^G_{f_1,f_2}=-\eta^G_{f_1,f_2}$) and for a star of mutually
  avoiding sets of SAWs and RWs
  ($\lambda^U_{f_1,f_2}=-\eta^U_{f_1,f_2}+f_1\eta^U_{2,0}$) read:
\begin{eqnarray} \nonumber
&&\lambda^G_{f_1,f_2}(\varepsilon)=\frac{ f_1f_2\varepsilon}{2}
\{1-( f_2 - 3 + f_1)\frac{\varepsilon}{4}[1-
(f_1+ \\ \label{2}  &&
f_2+ 3\zeta(3)-3)\frac{\varepsilon}{2}]\},
\\ \nonumber
&&\lambda^U_{f_1,f_2}(\varepsilon)=
-f_1 (3- f_1- 3 f_2)\frac {\varepsilon}{8} -
f_1 (43-33 f_1+  \\ \nonumber &&
8 f_1^2-
91 f_2+
42f_1 f_2+
18f_2^2) \frac {\varepsilon^2}{256}-
f_1 [675-
969f_1+  \\ \nonumber &&
456{ f_1}^{2}-
64{ f_1}^{3}-2463 f_2+
2290 f_1 f_2-
492{ f_1}^{2} f_2+ \\ \nonumber &&
1050{ f_2}^{2}-
504 f_1{ f_2}^{2}-
108{ f_2}^{3}-
\zeta (3)(712-
936 f_1 +  \\ \label{3} &&
224{ f_1}^{2}-
2652 f_2+
1188 f_1 f_2+
540{ f_2}^{2}) ]
\frac {\varepsilon^3}{4096},
\end{eqnarray}
where $\zeta(3)\simeq 1.202$ is the value of the Riemann zeta
function. While we do not display the exponents calculated at
fixed space dimension $D$ here explicitly, we show their numerical
values below.

The perturbative formulas of the kind given in Eqs.
(\ref{2})--(\ref{3}) may be evaluated for $D=3$ where the topological
complexity of the problem does not allow for an exact treatment. One
should be aware that for similar reasons an exact expression for the
partition function $Z_*$, formula (\ref{1}), can not be obtained for
$D=2$ either. However, appropriate identification of the
universality classes of $D=2$ critical behavior with conformally
invariant theories often allows to extract the exact values of
critical exponents.  In Refs.
\cite{Duplantier99} this was successfully performed for the $D=2$
copolymer star of two or more mutually avoiding bunches of SAWs and
RWs. Here, we analyze a copolymer star of one or two RWs or SAWs of
species 1 and a bunch of $f_2$ RWs of species 2.  In the notation of
the above equations the following exact expressions for the
corresponding exponents are derived from Refs.  \cite{Duplantier99}:
\begin{eqnarray}\label{4}
\lambda^G_{1,f_2}&=&\frac{4f_2+1+a}{8},\,\,\,
\lambda^G_{2,f_2} =\frac{12f_2+11+5a}{24},
\\ \label{5}
\lambda^U_{1,f_2}&=&\frac{12f_2-13+2a}{24},\,\,
\lambda^U_{2,f_2} =\frac{12f_2-16+5a}{24},
\end{eqnarray}
with $a=\sqrt{24f_2+1}$.

As far as this is not an exact solution in terms of the original
theory but rather an ``exact conjecture'' for the exponents we
performed one more check of the scaling exponents by a MC simulation
for the simplest case when the copolymer star consists of RWs only
($\lambda^G_{1,f_2},\lambda^G_{2,f_2}$). In these simulations the
stars are grown on a square lattice until a non-allowed intersection
occurs.  The number of stars $C(N)$ generated during the growing
process is accumulated for all chain lengths $N$. Exponents are
extracted assuming a power law $C(N)\sim N^{-\nu_0\lambda_{f_1,f_2}}$
where the RW correlation length exponent is $\nu_0=1/2$.  We generate
stars up to a maximum chain length $N_{\rm max}=10^3$ and accumulate
configurations until $C(N_{\rm max})=10^5$.  The number of successful
attempts to grow stars with longer chains decreases rapidly for higher
$f_1+f_2$ which increases simulation time drastically.  While for the
case $1+1$ a total of $10^6$ attempts were needed, this number rose to
$5\cdot10^9$ for $5+1$.  Therefore, we report results only for
$f_1+f_2\leq 6$.  From the bare simulation data we extracted the
numerical values of the exponents by an extrapolation to $1/N\to 0$.
In Table \ref{tab1} we report the results and the statistical error of
this extrapolation in the lines marked 'MC' and '$\pm\Delta$'. 
We compare with the exact numbers according to Eqs. (\ref{4}).  The
results are in fair agreement with a slightly growing discrepancy as
the number of RWs $f_2$ increases.

\begin{figure}
\begin{centering}
\includegraphics[width=80mm]{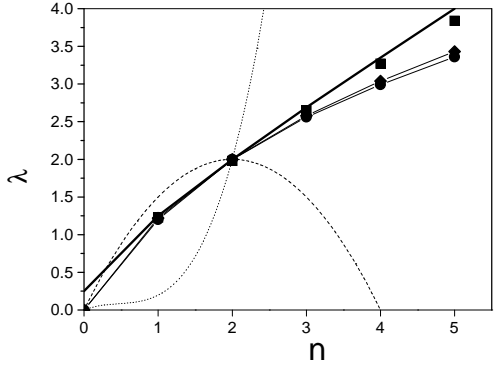}
\end{centering}
\caption{\label{fig1} Exponent $\lambda_{1,n}^G$. 
  Lines: exact (bold), bare $\varepsilon^2$ (dashed), and
  $\varepsilon^3$ (dotted) data. Symbols: MC ($\blacksquare$, size is
  larger than errorbars); the resummed $\varepsilon^3$ ({\large$\bullet$}),
  and $D=2$ data ($\blacklozenge$) are connected by thin lines.}
\end{figure}

In Fig. \ref{fig1} we compare the results with the perturbative data
of the $\varepsilon$-expansion naively adding successive orders of the
perturbation theory series (\ref{2}). We show the $\varepsilon^2-$ and
$\varepsilon^3-$approximations.  The values predicted by successive
orders of the $\varepsilon$-expansion for $\varepsilon=2$ do not seem
to correlate with each other nor with the exact values of the $d=2$
exponents. It is not only the large expansion parameter
$\varepsilon=2$ that spoils the convergence of the perturbation
theory. As is well known, the perturbation theory series of
renormalized field theory are asymptotic at best \cite{rgbooks} and
have zero radius of convergence \cite{Hardy}. An appropriate
resummation technique must be used to extract reliable data from the
series. Here we use a Borel resummation refined by a conformal mapping
\cite{LeGuillou80} as described for this particular case in Ref.
\cite{FerHol97}.

The resummation procedure allows us to restore convergence of the
$\varepsilon$-expansion and enables us to extract reliable exponent
values. These are also shown in Fig. \ref{fig1}.  The same resummation
is applied for the renormalization group expansions at fixed $D=2$
\cite{note2}. All data is summarized in Table \ref{tab1}. For
$\lambda^{G}_{1,2}$ perturbation theory gives the exact result at
first order with all higher orders vanishing \cite{Semenov88}. For
completeness we have included here as in Table \ref{tab2} the
corresponding $D=3$ dimensional results from \cite{FerHol97}.

Table \ref{tab2} contains the data for the resummed values of the
exponents $\lambda^U_{2,f_2}$ for the combined SAW and RW copolymer
stars.  We show the values obtained by the fixed $D=2$ technique and
the $\varepsilon$-expansion in comparison with their exact
counterparts (\ref{5}).  Again, the results are in fair agreement for
not too large numbers of chains $f_2$. Indeed as one can see from the
formulas (\ref{2})--(\ref{3}) the increase of $f_2$ leads to an
increase of the $\varepsilon$-expansion coefficients resulting in a
growing inaccuracy even of the resummed series.

In the perturbative treatment (which starts from the upper critical
dimension $D=4$ \cite{polymerRG}) the order of the chains in the star
does not matter for the scaling laws: there is always a possibility
for every chain to interact with any other chain constituting a star.
On the other hand, in $D=2$ the order of chains does matter: each
chain of the star will interact only with its direct neighbors.  This
topological restriction has to be taken into account when
comparing exact and perturbative results for $D=2$ linked polymers.

The data of Tables \ref{tab1}, \ref{tab2} convinces us that the
perturbation theory series for low numbers of chains $f_2$ is reliable
even for $D=2$. Also, we note that the $D=2$ copolymers with
$f_1=2$ are worse described by the perturbation theory than those
with $f_1= 1$ where the topological restriction is not present, due to
the symmetry of ordering.

\paragraph{Multifractal Spectrum}\label{III}
It is of special interest to notice another physical interpretation
for the copolymer star exponents (\ref{1}). For the diffusion of
freely moving particles in the presence of a polymer absorber it may
be shown \cite{Cates87} that the moments of particle density display
universal scaling laws in the vicinity of the absorbing polymer. In
particular, the $n$th moment $\langle\rho^n\rangle$ at distance $r$
from the core of an absorbing polymer star with $m$ chains scales as
\cite{Cates87,FerHol97}:
\begin{equation}\label{6}
<\rho^n(r)> \sim (R/r)^{-\lambda_{m,n}},
\end{equation}
here $<\dots>$ denotes the ensemble average over the configurations of
the absorbing polymers while otherwise we use the notations of formula
(\ref{1}). The harmonic measure that corresponds to this absorption
phenomenon possesses multifractal \cite{Cates87,Hentschel83,Halsey86}
properties: The flux $\phi$ of diffusing particles onto the polymer
star is interpreted as a measure defined on the fractal structure of
the star polymer.  This flux is proportional to the particle density
(\ref{6}) some small distance $\ell$ away from an absorbing point on
the star polymer.  The following scaling law for the
normalized moments of the harmonic measure at the core of the star
polymer is derived:
\begin{equation}\label{7}
<\phi^n>/<\phi>^n \sim (R/\ell)^{-\tau_{m,n}}.
\end{equation}
Comparing formulas (\ref{6}) and (\ref{7}) the obvious relation
between the exponents $\tau$ and $\lambda$ is:
\begin{equation}\label{8}
\tau_{m,n}= \lambda_{m,n} - n\lambda_{m,1}.
\end{equation}
Note, that here we use the standard definition \cite{Cates87} for the
spectrum of the exponents $\tau$, including normalization of the $n$th
moment of the flux in (\ref{7}) by $<\phi>^n$. The expression
(\ref{7}) is a multifractal measure as far as the spectrum $\tau$ is
non-trivial ($\tau\neq0$). Obviously, the spectrum $\lambda$ is
non-trivial as well, as one can see already from the second order
perturbative results (\ref{2}), (\ref{3}): $\lambda_{mn}\neq n
\lambda_{m1}$.

\begin{figure}
\begin{centering}
\includegraphics[width=80mm]{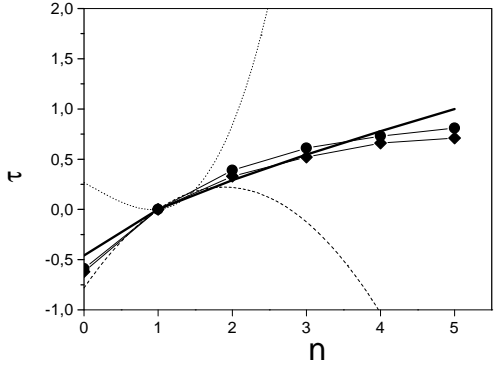}
\end{centering}
\caption{\label{fig2} The spectrum of exponents $\tau^U_{1n}$ for
the absorption at the end of a SAW. Symbols as in Fig.
\ref{fig1}.}
\end{figure}

Using the relation (\ref{8}) we now compare the values of the
exponents $\tau$ as obtained in the perturbative approach with those
of the MC simulations and the exact calculations. If the absorbing
polymer star is chosen to be RW-like, then the spectra $\tau$ are
calculated by substituting the exponents $\lambda^G$ into (\ref{8})
(and exponents $\lambda^U$ for SAWs, correspondingly). 
We show in Fig.
\ref{fig2} the resulting spectra for the
spectrum of the exponents $\tau^U_{1n}$ for the absorption at the
end-point of a SAW.  Again, we display the exact results derived from
formulas (\ref{4}) in comparison with the perturbative data with and
without resummation. We observe a similar behavior as for the spectra
$\lambda$: for low enough values of $n$ the resummed perturbative
renormalization group results are in reasonable agreement with the
exact predictions.

The description of multifractal phenomena often uses the spectral
function formalism \cite{Halsey86}. To obtain this function for the
absorption process on the center of a star of $m$ chains we analytically
continue the set of exponents $\tau_{mn}$ in the variable $n$ and calculate
the following Legendre transform
\begin{eqnarray} \label{9}
f_m(\alpha_{mn}) &=& -\tau_{mn} +n\alpha_{mn} + d (1-n),
\\ \nonumber
\alpha_{mn}&=&\frac{{\rm d}\tau_{mn}}{{\rm d}n}+d.
\end{eqnarray}
According to the standard definition \cite{Cates87} we have
included into (\ref{9}) the fractal dimension $d$ of the absorber.
In particular, this gives the maximal value of the spectral function
to be equal to $d-\tau_{m0}\equiv d-\lambda_{m0}$. There appears
to be no natural generalization of $d$ for arbitrary $m$. In our
presentation we use that the fractal dimension of a polymer star
is equal to that of a polymer chain ($d=2$ for a RW and $d=4/3$
for a SAW in $D=2$, correspondingly). Another way to define the
multifractal spectrum is chosen in Refs. \cite{Duplantier99}
where the $n$th moments (\ref{7}) are normalized not by
$<\phi>^n$ but by $<\phi>$. This choice has the remarkable feature
that without introducing an additional dimension $d$ in (\ref{9})
the maximum of $f(\alpha)$ $d=4/3$ for both the SAW and the RW absorbers
with $f_1=2$. However, this definition is not general, as already in
$D=3$ it does not lead to the correct location of the spectral
function maximum.

\begin{figure}
\begin{centering}
\includegraphics[width=80mm]{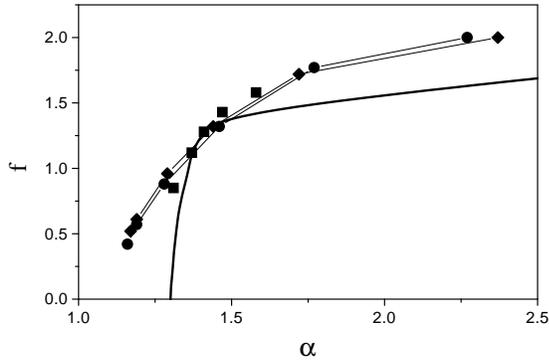}
\end{centering}
\caption{\label{fig3} The spectral function for the absorption at
a RW end. Symbols as in Fig. \ref{fig1}.}
\end{figure}

We display the spectral function for
the case of the absorption at a RW end in Fig.~\ref{fig3}. The
spectral function is negative for some values of $\alpha$. This
behavior was observed already in the perturbative studies of Ref.
\cite{Cates87} and related to the ensemble average in formulas
(\ref{6}), (\ref{7}) in contrast to the usual site average over the
support of the measure.  Comparing the perturbative, the MC and the
exact results we note that the region of $1<n<5$ in which reasonable
agreement for the spectra $\lambda$ and $\tau$ is found corresponds to
a rather small region in terms of the parameter $\alpha$ in the left
wing of the spectral functions.
\paragraph{Conclusions}\label{IV}
The field theoretical description of a polymer network of different
species has been derived recently \cite{FerHol97,costars} in the form
of renormalization group perturbation theory series. In this paper, we
compared and verified the perturbation expansions for $D=2$ with a
subsequently published exact study \cite{Duplantier99} and MC
simulations. We found that the series when appropriately resummed give
a reliable description of the scaling of copolymer stars with not too
many chains even for the $D=2$ dimensional case. Not only the increase
of the expansion parameter forbids application of the perturbative
results for higher numbers of chains: in $D=2$ topological
restrictions appear that are not taken into account by field theory.

We have analyzed the multifractal properties of the resulting spectra
of exponents and found that the spectrum $\tau(n)$ is in much better
coincidence with the exact counterpart than the spectral function
$f(\alpha)$. In the latter case the perturbative approach gives
comparable results 
only for a narrow region in the left wing of the
spectral function. Note however, that the right wing corresponds to
negative values of $n$, where obviously the analytic continuation of
the perturbative expansions as well as of the exact results is
speculative.

\begin{acknowledgments} 
Support of the Deutsche Forschungsgemeinschaft
is gratefully acknowledged.
\end{acknowledgments}

\begin{table}[!htp]
\begin{center}
\begin{tabular}{|c|lllllllll|}
\hline
 $f_1;\,f_2$ & 1;1  & 1;2  & 1;3  & 1;4  & 1;5  & 2;2  & 2;3 & 2;4 & 2;5 \\
\hline
exact        &1.25  & 2    & 2.693& 3.356& 4    & 2.916& 3.738& 4.510& 5.25\\
\hline
MC           & 1.251& 1.986& 2.662& 3.295& 3.908& 2.913& 3.703& 4.506& -- \\
$\pm\Delta$ 
           & 0.004& 0.004& 0.005& 0.015& 0.022& 0.005& 0.030& 0.039& -- \\
\hline
$D=2$    & 1.22 & 2.00 & 2.58 & 3.04 & 3.43 & 3.45 & 4.59& 5.52&6.34  \\
$\varepsilon^3$
             & 1.20  & 2.00 & 2.56 & 2.99 & 3.36 & 3.41 & 4.49& 5.37& 6.13 \\
\hline\hline
$D=3$         & 0.58 & 1.00 & 1.35 & 1.69 & 1.98 & 1.81 & 2.53 & 3.17 & 3.75\\ 
$\varepsilon^3$& 0.56& 1.00 & 1.33 & 1.63 & 1.88 & 1.77 & 2.45 & 3.01 & 3.51\\
\hline
\end{tabular}
\caption
{\label{tab1} Exponents $\lambda_{f_1,f_2}^G$ ($f_1$ RWs and $f_2$ RWs)
calculated by different techniques. Lower part: $D=3$ results.}
\end{center}
\end{table}
\begin{table}[!htp]
\begin{center}
\begin{tabular}{|c|llllllllll|}
\hline
$f_1;\,f_2$ & 1;1 & 1;2  & 1;3  & 1;4  & 1;5 & 2;1 & 2;2  & 2;3 & 2;4 & 2;5 \\
\hline
exact       & .38 & 1.04 & 1.67 & 2.28 & 2.88&.88  &1.79  &2.61 &3.39 &4.13 \\
\hline
$D=2$
& 0.34 & 1.01 & 1.54 & 2.02 & 2.41 & 0.72 & 1.85 & 2.80 & 3.64 & 4.39 \\
$\varepsilon^3$
& 0.32 & 1.03 & 1.57 & 2.01 & 2.41 & 0.76 & 1.87 & 2.82 & 3.62 & 4.34 \\
\hline\hline
$D=3$ 
& 0.15 & 0.51 & 0.81 & 1.07 & 1.32 & 0.42 & 1.02 & 1.57 & 2.05 & 2.49 \\
$\varepsilon^3$ 
& 0.17 & 0.53 & 0.81 & 1.09 & 1.36 & 0.42 & 1.04 & 1.63 & 2.15 & 2.65 \\
\hline
\end{tabular}
\caption
{\label{tab2} Exponents $\lambda_{f_1,f_2}^U$ ($f_1$ SAWs and $f_2$
  RWs) calculated by different techniques. Lower part: $D=3$ results.}
\end{center}
\end{table}

\begin{thebibliography}{22}
\bibitem{note2} In this case the expansion is performed in a
``pseudo-$\varepsilon$'' parameter, as explained e.g. in:
\cite{LeGuillou80}.
\end{thebibliography}
\end{document}